\documentclass[prl,aps,twocolumn,showpacs]{revtex4}
\usepackage{graphicx}

\begin{document}
\title{Enhancement of Macroscopic Quantum Tunneling by Landau--Zener
Transitions}

\author{Joachim Ankerhold} 
\email{ankerhold@physik.uni-freiburg.de}
\author{Hermann Grabert}
\affiliation{
Physikalisches Institut, Albert-Ludwigs-Universit{\"a}t
Freiburg, Hermann-Herder-Stra{\ss}e 3, D-79104 Freiburg, Germany}

\date{\today}
          

\begin{abstract}
Motivated by recent realizations of qubits with a readout by
macroscopic quantum tunneling in a Josephson junction, we study
the problem of barrier penetration in presence of coupling to a 
spin-${1\over 2}$ system. It is shown that when the diabatic
potentials for fixed spin intersect in the barrier region,
Landau--Zener transitions lead to an enhancement of the tunneling
rate. The effect of these spin flips in imaginary time is
in qualitative agreement with experimental observations.
\end{abstract}

\maketitle

Macroscopic Quantum Tunneling (MQT) in Josephson systems \cite{leggett}
has been studied in detail both experimentally \cite{devoret} and
theoretically \cite{grabert} in the eighties but has gained
renewed interest very recently \cite{vion,martinis} since the
exponential dependence of the MQT rate on parameters allows for a
high fidelity readout of qubits based on superconducting circuits.
In this context a new variety of quantum tunneling problem arises,
namely, barrier penetration in presence of coupling to a 
spin-${1\over 2}$ describing the qubit. For fixed spin
the phase of the Josephson junction tunnels through a diabatic
potential barrier, and the exponentially large difference 
between the MQT rates corresponding to the barrier
potentials for the two spin directions is exploited in the 
readout to determine the spin state. 
However, there is an interesting parameter range where
the diabatic potentials intersect in the barrier region so that
Landau-Zener (LZ) transitions may arise. In contrast to the well-studied
problem of LZ transitions in real time \cite{LZ}, here, within the standard
semiclassical approach to quantum tunneling, these transitions
occur in ``imaginary" time. It is shown that they cause an enhancement
of the tunneling rate in the appropriate parameter range
in agreement with experimental observations \cite{vion2}.\hfill\break
\vspace*{0.1cm}
To motivate the Hamiltonian underlying this work, we briefly consider the 
quantronium circuit \cite{vion} consisting of a small superconducting 
electrode, the island, carrying an excess number of $N$ Cooper pairs. 
This island is attached to a superconducting loop via two Josephson junctions 
with Josephson energy $E_J/2$, and it can be biased by a voltage source $U$
through a gate capacitance $C_g$ inducing the dimensionless gate
charge $N_g= C_gU/2e$. For sufficiently large Coulomb charging energy
$E_C= (2e)^2/2C$, where $C$ is the island capacitance, and for $N_g$
near ${1\over 2}$ only two charge states $N=0,1$ are relevant, and the island
corresponds to an effective spin-${1\over 2}$ system that can serve as a 
qubit manipulated via the gate. For readout purposes the superconducting loop
is interrupted by a Josephson junction with Josephson energy $E_J'$ that is
shunted by a large capacitance $C'$ and can be biased by a current $I_b$.
This circuit is described by the Hamiltonian 
\begin{eqnarray}
{\cal H}&=& E_C (N-N_g)^2- E_J\ \cos\left(\frac{\theta+\phi}{2}\right)\,
\cos(\delta)\nonumber\\
&+&\frac{Q^2}{2 C'}-E_J'\, \cos(\theta)-
\frac{\hbar I_b}{2e} \theta
\nonumber
\end{eqnarray}
with the conjugate observables $[N,\delta]=-i$ for the qubit and
$[Q,\theta]= -2i e$ for the readout junction, respectively. 
$\phi$ denotes the external magnetic flux through the superconducting
loop in units of the flux quantum $\hbar/2e$. 

Within the subspace spanned by the eigenvectors $|0\rangle,
|1\rangle$ of the operator $N$, the qubit can be described by Pauli matrices 
 $\sigma_i$. Further, measuring all energies in units of $E_J'$, we
obtain the dimensionless Hamiltonian
\begin{equation}
H = \epsilon \, \sigma_z-j \,
\cos\left(\frac{\theta+\phi}{2}\right)\, \sigma_x +\frac{p_\theta^2}{2m } -
\cos(\theta) -i_b\, \theta\label{ham3}
\end{equation}
with the dimensionless parameters  
$\epsilon=(E_C/E_J')(N_g-{1\over 2})$, $j=E_J/E_J'$, and 
$i_b= \hbar I_b/2eE_J'$. The variables $p_\theta=Q/2e$ and $\theta$, with the
commutator $[p_\theta , \theta ] = -i$, may be viewed 
as dimensionless momentum and coordinate of a particle with 
dimensionless  mass $m= C'E_J'/4e^2$. 
We do not discuss here the manipulations of the qubit done for
$i_b=0$, but address the readout when the bias current is
increased to a value slightly below the dimensionless critical
current $i_c= 1$. The problem at hand then is tunneling
of this ficticious particle through a barrier of the  potential 
$\cos(\theta) -i_b\theta$ in presence of the interaction with the spin.

In view of the large mass $m$ the coordinate $\theta$ is almost
a classical variable. When the kinetic energy $p_\theta^2/2m$ is
neglected, the Hamiltonian (\ref{ham3}) can easily be diagonalized with
the eigenvalues
\[
\lambda_\pm (i_b,\theta) = -\cos(\theta)-i_b \theta \pm
\sqrt{\epsilon^2+j^2\,
\cos^2[(\theta+\phi)/2]} \nonumber
\]
that determine two adiabatic potential surfaces. At zero temperature
and for vanishing bias current the system will approach the minimum 
of the lower surface $\lambda_- (i_b=0,\theta)$ 
which for small $\epsilon$ and $j$ lies  close to $\theta=0$.
As usual in MQT experiments, the switching of the bias current 
form $0$ to a value close to $i_c=1$ is slow compared to the 
characteristic time scales of the circuit. When the system follows 
the bias current adiabatically, the particle lies at finite bias $i_b$
near the minimum $\theta_-(i_b)$ of the adiabatic potential 
$\lambda_-(i_b,\theta)$. 
This state serves as the initial state for the calculation 
of the tunneling rate.

For this initial state it is natural to use the spin eigenvectors 
$|\theta_-,+\rangle,|\theta_-,-\rangle$
associated with the eigenvalues $\lambda_\pm(i_b,\theta_-)$ as a
basis for a matrix representation of the Hamiltonian (\ref{ham3}).
We then find
\begin{equation}
H=
\left(\begin{array}{cc}
{\displaystyle \frac{p_\theta^2}{2m}}+V_+(\theta)\ \  &\ \  \Delta(\theta)\\
\Delta(\theta)\ \  &\ \  {\displaystyle\frac{p_\theta^2}{2m}}+V_-(\theta)
\end{array}\right)
\label{matrix}
\end{equation}
where
\begin{equation}
V_\pm(\theta)=-\cos(\theta)-i_b\, \theta\pm\left(
\sqrt{\epsilon^2+V_0^2}\, +\, \kappa_0\,[V(\theta)-V_0]\right)
\label{potpm}
\end{equation}
are now two diabatic surfaces corresponding to the two spin 
orientations $|\theta_-,+\rangle,|\theta_-,-\rangle$, and
\[
\Delta(\theta)=\epsilon
\kappa_0  \left[1-\frac{V(\theta)}{V_0}\right]
\]
is the $\theta$ dependent coupling between them.
For convenience, we introduced 
\[
V(\theta)=j \cos[(\theta+\phi)/2]\, ,
\]
and $V_0=V(\theta_-)$, as well as 
\[
\kappa_0=2 V_0
\frac{\epsilon+\sqrt{\epsilon^2+V_0^2}}{(\epsilon+\sqrt{\epsilon^2+V_0^2})^2
+V_0^2}\, .
\]
Apparently, for $\theta=\theta_-$ the Hamiltonian (\ref{matrix})
is diagonal. It further becomes diagonal in the limits $\epsilon\to 0$ 
or $j \to 0$. 

Depending on the external flux $\phi$ and $\epsilon/j$ the two 
diabatic potentials (\ref{potpm}) may intersect.
As can be seen from Fig.~1 this is always the case for
$\phi$ near $\pi\over 2$ and $\epsilon$ small compared to $j$.
\begin{figure}
\center
\vspace*{-1.8cm}
\includegraphics[height=12cm,draft=false]{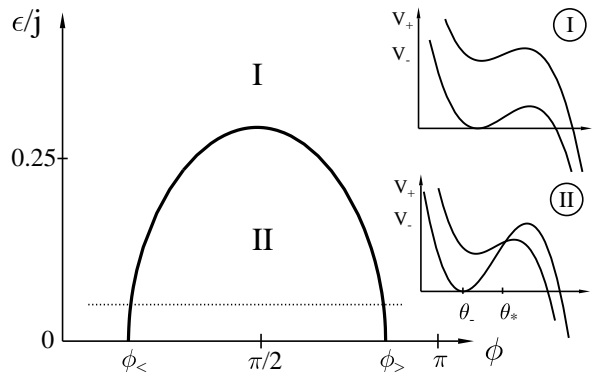}
\label{fig1}
\vspace*{-5.5cm}
\caption{Diabatic potential surfaces (\ref{potpm}) in dependence on 
$\epsilon/j$ and $\phi$ for a bias current of $i_b=0.96$. In the region 
denoted by {\bf I} the potentials do not intersect as shown in the upper inset 
displaying the potentials for $\epsilon/j=0.4 , \phi =0.5$. 
In region {\bf II} the potentials cross in the barrier region as shown in 
the lower inset for $\epsilon/j=0.1 , \phi =1.5$. The phases $\theta_-$
in the metastable initial state before tunneling and $\theta_*$
at the crossing point are also indicated. The dotted line refers to Fig.~3.}
\end{figure}
Near such a crossing point $\theta_*$, which is determined by 
\[
\frac{\epsilon}{j}=\left[-\cos\left(\frac{\theta_-+\phi}{2}\right)\
\cos\left(\frac{\theta_*+\phi}{2}\right)\right]^{1/2}\, ,
\]
the diabatic potential surfaces are strongly
coupled by the off-diagonal element of the Hamiltonian (\ref{matrix}).
To quantify the strength of this coupling we may introduce the parameter
\begin{equation}
g(\theta)=\left|\frac{\Delta(\theta)}{V_+(\theta)-V_-(\theta)}\right|
\nonumber\label{para}
\end{equation}
which diverges at the crossing point $\theta_*$. A LZ region 
where spin-flips may occur can then be defined by the condition 
$g(\theta) > 1$. We assume that this region is restricted to the
vicinity of $\theta_*$. In fact, when the bias current is sufficiently
far from 
the critical current to allow for several states in the metastable 
minimum, the LZ region turns out to be narrow except near the boundary 
between the regions {\bf I} and {\bf II} in Fig.~1. This crossover region will
be discussed further below.

To determine the tunneling rate we employ the ``bounce technique" 
\cite{bounce} which relates the rate $\Gamma$ essentially to
an imaginary time trajectory in the inverted potential, the 
so--called bounce. This method is equivalent to WKB and starts out 
form the partition function of the metastable system
\[
Z={\rm Tr}\left\{ |\theta_-,-\rangle \langle \theta_-,-|\ {\rm
e}^{-\beta H}\right\}
\]
which has to be evaluated in the semiclassical limit for 
$\beta \rightarrow \infty$.
Within the path integral representation this takes
the form
\begin{eqnarray}
Z &=&\int {\cal D}[\theta]\ {\rm e}^{-S_-[\theta]} \Bigg\{ 1+ 
\nonumber\\
&& +\sum_{n=1}^\infty\, \int_0^{\beta}ds_{2n} \cdots
\int_0^{s_2}ds_1\, \Delta[\theta(s_{2n})] \cdots
\Delta[\theta(s_{1})]\nonumber \\ 
&&   \times\, \exp\left[
\sum_{k=1}^n\, \int_{s_{2k-1}}^{s_{2k}} d\tau
(V_-[\theta]-V_+[\theta])\right]\Bigg\}\label{part1}\nonumber
\end{eqnarray}
where the path sum runs over all orbits with period $\beta$
switching $2n$ times between $V_-$ and $V_+$ at times $s_1<s_2<\ldots
<s_{2n}$. The Euclidian action on $V_-$ is
\[ S_-=\int_0^{\beta}
d\tau \left[\frac{p_\theta^2}{2m}+V_-(\theta)\right].
\]
In the semiclassical limit $Z$  decomposes into $Z_{sc}\approx
Z_w+Z_0+Z_2$. Here $Z_w$ is the partition function of the well which
is obtained by summing over paths in the vicinity of the 
trivial trajectory $\theta(\tau) = \theta_-$ sitting at the well minimum.
$Z_0$ is the contribution of paths in the vicinity of the standard non--flip 
bounce trajectory in $V_-$. In region {\bf II} we also have to take
into account the contribution $Z_2$ from paths that flip when the
bounce traverses the LZ region. Trajectories with four and more spin--flips
can be neglected away form the boundary between regions {\bf I} and {\bf II}.
Both the bounce and spin--flip bounce are saddlepoint trajectories with
an unstable fluctuation mode which after an  analytical continuation 
\cite{bounce} yield imaginary and, compared to $Z_w$, exponentially 
small contributions that determine the rate. 
Following standard procedures, we obtain for the dimensionless
rate, in units of $E_J'/\hbar$,
\begin{equation}
\Gamma= \lim_{\beta\to\infty} \frac{2}{\beta} \Im(Z_0+Z_2)/Z_w 
= \Gamma_0+\Gamma_2.\label{totalrate}
\end{equation}

To evaluate this explicitly, we first note that within the barrier region 
the potentials $V_\pm$ can very accurately be approximated by cubic
polynomials. It is convenient to introduce for each diabatic potential
the frequency at the well bottom
\[
\omega_\pm^2=V_\pm''(\theta_\pm)/m\ ,
\]
and a scaled distance 
\[
x_*^\pm=(\theta_*-\theta_\pm)/(\theta_\pm^0-\theta_\pm)
\]
between the well bottom and the ``exit point" $\theta_\pm^0$ where
$V_\pm(\theta_\pm^0)=V_\pm(\theta_\pm)$. Further,
$V_b^\pm$ is the barrier height with respect to the minimum 
$V_\pm(\theta_\pm)$.

Since we are interested in the limit $\beta \to \infty$, it
is natural to look for solutions in the time interval 
$s\in [-\beta/2,\beta/2]$. The  simple bounce trajectory in the 
inverted potential $-V_-(\theta)$ then reads (see Fig.~2)
\[
\theta_{\rm bounce}(s)=\theta_-+(\theta_-^0-\theta_-)/{\rm
cosh}^2(\omega_-\, s/2)  . 
\]
This trajectory dominates the non-flip contribution $Z_0$ that
has been evaluated previously yielding the well-known MQT rate in 
the absence of damping
\cite{caldeira}
\begin{equation}
\Gamma_0= 6\, \sqrt{6\, \omega_-V_b^-/\pi}\
\exp\left(-\frac{36}{5}\frac{V_b^-}{\omega_-} 
\right)\label{simplerate}
\end{equation}
which determines the rate in region {\bf I} of Fig.~1.

In region {\bf II} the semiclassical trajectory may
switch to the potential surface $V_+(\theta)$ in the LZ region.
The lowest order flip contribution reads
\begin{eqnarray}
Z_2 &=&\int {\cal D}[\theta]\ {\rm e}^{S_-[\theta]}\
\int_0^{\beta} ds_{2} \int_0^{s_2} ds_1 \Delta[\theta(s_{2})]
\Delta[\theta(s_{1})]\nonumber\\
&&\hspace{1cm}\times\, \exp\left[
\int_{s_{1}}^{s_{2}} d\tau
(V_-[\theta]-V_+[\theta])\right].\label{part2}\nonumber
\end{eqnarray}
To determine $Z_2$ we proceed as follows:
First, the action for the flip bounce is calculated for arbitrary
flipping times $s_1 < s_2$. Due to energy conservation and the
periodic boundary condition one finds that the restriction
$\theta(s_1)=\theta(s_2)$ applies. For a path running in the interval
$[-\beta/2,\beta/2]$ this means that the flips have to occur symmetrically
around $s=0$. 
Second, as a function of $s_1$ the action has a minimum at an
optimal flipping time $s_1^*$ determined by
\begin{equation}
p_\theta\,\left. \frac{\partial t_+[\theta]}{\partial
\theta}\
\{V_+[\theta]-V_-[\theta]\}\right|_{s=s_1^*}=0\label{extra} 
\end{equation}
where $t_+[\theta]$ is the time the bounce spends on the $V_+$
surface. As one might have guessed, Eq.~(\ref{extra}) yields 
$V_+[\theta]=V_-[\theta]$, so that the optimal flips occur at 
the intersection point $\theta_*$  of the diabatic potentials.
Then $\theta_*=\theta(s_1^*)$ and $s_2^*-s_1^*= t_+(\theta_*):=t_+$. 
[Other solutions of Eq.~(\ref{extra}) with $p_{\theta}(s_1^*)=0$
mean that  flips occur either in the well or at the turning
point. In both cases the orbit has no energy to run on $V_+$ and one
regains the simple bounce action on $V_-$. Further, solutions 
with  $\partial t_+[\theta]/\partial\theta=0$ corresponds to a 
maximum of the action.]

This way the trajectory at the saddlepoint of the action is obtained as
\[
\theta_{\rm flip}(s)= h(|s| - \frac{t_+}{2})\ \theta_{\rm flip}^-(s)
+h(\frac{t_+}{2}-|s|)\ \theta_{\rm flip}^+(s).
\]
Here the step functions $h(\cdot)$ select the time segments
spend on the two potential surfaces, where
\begin{equation}
t_+=\frac{2}{\omega_+\lambda}\ {\rm F}(\varphi|\bar{m}).\label{plustime}
\end{equation}
The parameter $\lambda = [P'(x_0)]^{1/4}$ is determined by the
slope $P'$ of the polynomial
\begin{equation}
P(x)=x^3-x^2-\frac{4}{27}\,\rho, \hbox{with\ } 
\rho=\frac{V_+(\theta_+)-V_-(\theta_-)}{V_b^+}\label{poly}
\end{equation}
at its zero where $P(x_0)=0$,  and 
${\rm F}(\varphi|\bar{m})$ is the elliptic integral of the first kind
with modulus $\bar{m}=1/2+P''(x_0)/8\lambda^2$ and angle
$\cos(\varphi)=[\lambda^2-(x_0-x_*^+)]/[\lambda^2+(x_0-x_*^+)]$. 
Further, 
$\theta_{\rm flip}^-(s)=\theta_{\rm bounce}[s-{\rm sign}(s) \alpha]$ 
describes the segments of the flip bounce on the surface $V_-$ 
where it coincides with the simple bounce apart from a phase 
\[
\alpha=\frac{t_+}{2}-\frac{2}{\omega_-}\, {\rm Arccosh}(\sqrt{x_*^-}) .
\]
Finally, the segment running on $V_+$ follows as
\[
\theta_{\rm
flip}^+(s)=\theta_++(\theta_+^0-\theta_+)\left[x_0+\lambda^2\frac{{\rm 
cn}(\lambda s\omega_+|\bar{m})-1}{{\rm
cn}(\lambda s\omega_+|\bar{m})+1}\right]
\]
where
${\rm cn}(\cdot)$ is a Jacobi function (see Fig.~2). 
\begin{figure}
\center
\vspace*{-1.3cm}
\includegraphics[height=11cm,draft=false]{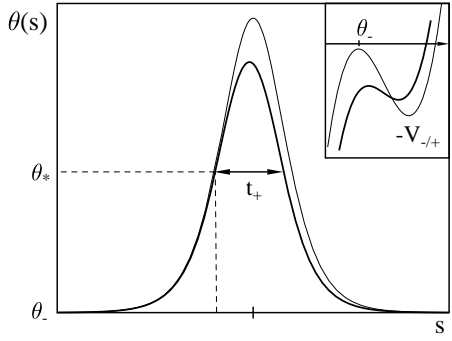}
\label{fig2}
\vspace*{-5.0cm}
\caption{Non-flip bounce (thin line) and flip bounce (thick line)
 vs.  time. The inset shows the inverted potential
 surfaces $-V_-$ (thin line) and $-V_+$ (thick line), see text for details.}
\end{figure}

For the action of this flip bounce one finds
\begin{eqnarray}
S_*&=&\frac{18\, V_b^-}{5\, \omega_-}\,
\left[\sqrt{(1-x_*^-)^3}\,
\left(3\sqrt{(1-x_*^-)}-5\right)+2\right]\nonumber\\
&+&\frac{18\, V_b^+}{5\, \omega_+}\bigg[\frac{2x_0-2\lambda^2+9\, \rho}{2\lambda}\, {\rm F}(\varphi|\bar{m})+2\lambda {\rm
E}(\varphi|\bar{m})\nonumber\\
&-& 
\frac{1}{2}\sqrt{-P(x_*^+)}\left(\frac{4}{\lambda^2+x_0-x_*^+}+P''(x_*^+)\right)\bigg].\nonumber
\end{eqnarray}
Here, E$(\cdot)$ is the elliptic integral of the second kind. 

Next, the action is expanded around  $S_*$ with respect
to variations of the flipping times up to second order. 
For this purpose it is convenient to introduce sum,
$u=s_1+s_2$, and difference, $v=s_2-s_1$, times, respectively. The
Gaussian factor describing deviations from $s_1^*, s_2^*$ then depends
only on $v$, i.e.\  only on the relative position
of the flips but not on the absolute position of the bounce in
time. Hence, like the conventional non--flip bounce, the flip bounce
has one zero mode (corresponding to the integration over $u$), 
while the integration over $v$ is weighted with a Gaussian factor 
$\exp[-\Omega_1^2/2\, (v- t_+[\theta_*])^2]$.
The frequency 
\[
\Omega_1=\frac{2}{m} V_-(\theta_*)\, [V_+'(\theta_*)-V_-'(\theta_*)]
\left.\frac{\partial t_+(\theta)}{\partial
\theta}\right|_{\theta=\theta_*}\label{omega}
\]
is proportional to the second derivative of
the action at $s_1^*, s_2^*$. Here, $t_+(\theta)$ for arbitrary
$\theta$ is given by the expression
(\ref{plustime}) 
where the parameters $\lambda(\theta)$ and $\bar{m}(\theta)$ are gained by
replacing in 
(\ref{poly}) 
$\rho\to
\rho(\theta)=\rho+[V_-(\theta)-V_+(\theta)]/V_b^+$ 
and the angle
$\varphi(\theta)$ 
follows from
$\cos(\varphi)=[\lambda^2-(x_0-x^+)]/[
\lambda^2 + (x_0-x^+)]$
with $x^+(\theta)=(\theta-\theta_+)/(\theta_+^0-\theta_+)$.

Finally, fluctuations in $\theta$ 
are calculated in the usual way \cite{bounce} from a ratio of two
determinants
where the zero eigenvalue is omitted and replaced by the proper
zero mode normalization factor while the unstable mode is accounted for
by an analytic continuation leading to the imaginary part.

Eventually, one obtains in the limit $\beta\to \infty$ for the
contribution $\Gamma_2$ to the rate (\ref{totalrate})
\[
\Gamma_2= 6\sqrt{\frac{3V_b^-}{\omega_-}}\, \frac{\Delta(\theta_*)^2}{
\Omega_1}\  {\rm erfc}\left[-\frac{ \Omega_1
\, t_+(\theta_*)}{\sqrt{2}}\right]  {\rm e}^{-S_*}.\label{rat2}
\]
This can be combined with (\ref{simplerate}) to yield the central
result of this paper, namely, the decay
rate in region {\bf II} 
\[
\Gamma=\Gamma_0\ \left\{1+\frac{\Delta(\theta_*)^2}{ \omega_-\,
\Omega_1}\ \sqrt{\frac{\pi}{2}}\ {\rm erfc}\left[-\frac{ \Omega_1
\, t_+(\theta_*)}{\sqrt{2}}\right] \ {\rm e}^{(S_--S_*)}\right\}
\label{total}
\]
with $S_-= \frac{36}{5} V_b^-/\omega_-$. The corresponding
rate enhancement is shown in Fig.~3. 
\begin{figure}
\center
\vspace*{-1.2cm}
\includegraphics[height=10cm,draft=false]{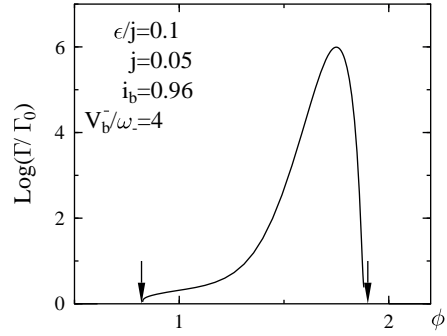}
\label{fig3}
\vspace*{-4.5cm}
\caption{Rate enhancement vs. magnetic flux
along the dotted line in Fig.~1. Arrows indicate the boundary between 
regions {\bf I} and {\bf II}.}  
\end{figure}
Apparently, there is a pronounced exponential increase of the total 
rate due to spin flips in the LZ range along the bounce. This is in
agreement with experimental findings of the peak current
variation with magnetic flux in the quantronium device \cite{vion2}. 

As one approaches the boundary between {\bf I} and {\bf II} the
LZ region grows and in a narrow boundary layer the parameter
$g(\theta)$ in (\ref{para}) is larger than 1 in the entire barrier
range. Then, multi--spin flips can occur anywhere along the bounce and
also during the switching on of the bias current. 
However, since the exponential factors
of $\Gamma_0$ and $\Gamma_2$ coincide at the boundary, 
this breakdown of the nonadiabatic approach utilized here, essentially
reduces to a prefactor effect smoothing the transition
between the results in region {\bf I} and {\bf II}. A more
interesting extension of the present work would consider initial
states where due to manipulations via the charge gate the system moves
on the upper adiabatic potential surface $\lambda_+(i_b,\theta)$.
This will be addressed in future work.

The authors would like to thank D.~Esteve, P.~Joyez,
H. Pothier, C.~Urbina, and D.~Vion for motivating this work
and fruitful discussions. Financial support has been provided by the
Deutsche Forschungsgemeinschaft (Bonn) and the European Community.


\begin{thebibliography}{10}

\bibitem{leggett}
A.J. Leggett, Prog. Theor. Phys. (Suppl.) {\bf 69}, 80 (1980).

\bibitem{devoret}
M.H. Devoret, D. Esteve, C. Urbina, J. Martinis, A. Cleland, and
J. Clarke, in {\em Quantum Tunnelling in Condensed Media}
edited by Yu. Kagan and A.J. Leggett (Elsevier, Amsterdam, 1992); and
references therein.

\bibitem{grabert}
H. Grabert, in {\em SQUID'85 - Superconducting Quantum Interference 
Devices and their Applications} edited by H.D. Hahlbohm and H. L{\"u}bbig
(Walter de Gruyter, Berlin, 1985); and references therein.

\bibitem{vion}
D. Vion, A. Aassime, A. Cottet, P. Joyez, H. Pothier, C. Urbina,
D. Esteve, and M.H. Devoret, Science {\bf 296}, 886 (2002).

\bibitem{martinis}
J.M. Martinis, S. Nam, J. Aumentado, and C. Urbina,
Phys. Rev. Lett. {\bf 89}, 117901 (2002).

\bibitem{LZ}
L.D Landau, Phys. Z. Sowjetunion {\bf 2}, 46 (1932);
C. Zener, Proc. R. Soc. London A {\bf 137}, 696 (1932)

\bibitem{vion2}
D. Vion (private communication).

\bibitem{bounce}
W.H. Miller, J. Chem. Phys. {\bf 62}, 1899 (1975);
M. Stone, Phys. Lett {\bf 67B}, 186 (1977);
C.G. Callan and S. Coleman, Phys. Rev. D. {\bf 16}, 1762 (1977).

\bibitem{caldeira}
A.O. Caldeira and A.J. Leggett, Ann. Phys, (N.Y.) {\bf 149}, 374 (1983).

\end{thebibliography}
\end{document}